\documentclass[twocolumn,amsmath,amssymb,showpacs,pra]{revtex4-1} 
\usepackage{graphicx}

\begin{document}

%%%%%%%%%%%%%%%%%% title page information %%%%%%%%%%%%%%%%%%
\title{Bistability and instability of dark-antidark solitons\\in the cubic-quintic nonlinear Schr\"odinger equation}

\author{M. Crosta$^{1,2}$}
%\address{Dept. of Physics, Sapienza University of Rome, I-00185, Rome, Italy}
\author{A. Fratalocchi$^2$}
%\address{PRIMALIGHT, Faculty of Electrical Engineering; Applied Mathematics and Computational Science, King Abdullah University of Science and Technology (KAUST), Thuwal 23955-6900, Saudi Arabia}
\homepage{http:www.primalight.org} %% author's URL, if desired
\email{andrea.fratalocchi@kaust.edu.sa} %% email address is required
\author{S. Trillo$^3$}
%\address{Dipartimento di Ingegneria, Universit\`{a} di Ferrara, Via Saragat 1, 44122 Ferrara, Italy}
\email{stefano.trillo@unife.it}

\affiliation{ 
$^1$Dept. of Physics, Sapienza University of Rome, 
I-00185, Rome, Italy\\
$^2$PRIMALIGHT, Faculty of Electrical Engineering; Applied Mathematics and Computational Science, King Abdullah University of Science and Technology (KAUST), Thuwal 23955-6900, Saudi Arabia\\
$^3$Dipartimento di Ingegneria, Universit\`{a} di Ferrara, Via Saragat 1, 
44122 Ferrara, Italy
} 

\date{\today} 

\begin{abstract}
We characterize the full family of soliton solutions sitting over a background plane wave and ruled by the cubic-quintic nonlinear Schr\"odinger equation
in the regime where a quintic focusing term represents a saturation of the cubic defocusing nonlinearity. 
We discuss existence and properties of solitons in terms of catastrophe theory and fully characterize bistability and instabilities of the 
dark-antidark pairs, revealing new mechanisms of decay of antidark solitons.
\end{abstract}
 % from new scheme 2010

\maketitle 

\section{Introduction}

Optical spatial solitons are important for their capability to beat diffraction and their potential for engineering a variety of optical reconfigurable structures including (and not limited to) couplers, deflectors and logic gates. In Kerr media where the paraxial propagation is described by the nonlinear Schr\"odinger (NLS) equation, only one soliton solution exists once the parameters (i.e., nonlinearity and peak intensity or width) are fixed and that solution is stable. 
However, more general nonlinear responses can result into bistability of solitons (strictly speaking solitary waves)
and/or their instability against the growth of weak perturbations.

In this paper we are interested to investigate such features for solitons sitting on a finite background (i.e., dark-like) in the context
of the Cubic-Quintic NLS (CQNLS) with a defocusing cubic and focusing quintic nonlinear response.
The importance of such model lies in the fact that it constitutes the simplest model for a defocusing saturable Kerr effect \cite{Krolikowski93,Kivshar98},
whose parameters can be effectively measured in a relatively simple way by two-wave coupling or Z-scan \cite{Wise05,Zhang07}.

As far as bistability is concerned, the case of dark solitons has been investigated with reference to various model
including the CQNLS \cite{Barashenkov88,Mulder89,Herrmann92,PhysRevA92.46.4185,Tanev97,Kivshar98}.
In particular Hermann has shown that dark solitons exhibit bistability of the second kind, i.e. 
characterized by solutions possessing the same full-width-half-maximum (FWHM), albeit possessing different amplitudes (and generally invariants of motion). This type of bistability was introduced in Refs. \cite{Gatz91,Eix93} by Gatz and Hermann, to distinguish it from the earlier definition
\cite{PhysRevLett.55.1291,PhysRevLett.57.778,PhysRevA.35.466} which implies the existence of different solutions possessing the same value of one invariant of motion (e.g., the power) for different values of the internal parameter, typically the nonlinear propagation constant $\beta$.
The analysis carried out by Herrmann, however, is limited to stationary solitons, while a full family of moving dark solitons can exist.
The analysis is further complicated by the fact that the CQNLS in the regime considered here is known
to possess coexisting antidark (or bright on pedestal) solutions \cite{Kivshar96,Tanev97}. 
The full family of dark and antidark solitons, once parametrized by the velocity, exhibits intriguing features which have been overlooked,
and which we discuss below. Furthermore, whether the full family of the dark-antidark moving pairs is stable or not, and which are the instability mechanisms is still an open problem. Our systematic investigation of these problems provides two answers: 
(i) it shows that the criterium demonstrated by Barashenkov \cite{PhysRevLett.77.1193} for dark solitons 
provides the correct exhaustive answer to the stability problem also for antidark solitons;
(ii) it clarifies that the decay of antidark solitons can follow new scenarios in proper regions of the parameter space,
rather than always blowing-up as conjectured in the previous literature, though collapse is in general allowed
even in 1+1D because of the high power of the focusing term.\\
\indent Besides being important {\em per se}, the knowledge of the dynamics of the whole soliton family of the CQNLS is also important
in view of recent studies which extend the investigation of competing nonlinearities to the nonparaxial \cite{PhysRevA.81.053831}  and nonlocal  \cite{PhysRevA.82.063829,ZhouJOSAB11} regimes. Moreover,  the full characterization of the soliton solutions
and their instabilities constitute the starting ground for describing the feature of dispersive shock waves (DSW, involving multiple solitons in the weakly dispersive regime) \cite{Hoefer06},
an active area of research where succesful experiments have been recently performed  in non-Kerr media  under different excitation conditions 
\cite{Wan07,PhysRevLett.102.083902,PhysRevA.83.053846}.
In this respect, here we provide the first prediction of a dispersive shock wave produced directly by the decay of a solitary wave of the CQNLS model.

\section{Dark-Antidark solutions}
We start from the (dimensionless) CQNLS equation:
\begin{equation} \label{nls35}
i  \frac{\partial u}{\partial z} +  \frac{1}{2}\frac{\partial^{2} u}{\partial x^{2}} -  \left| u \right|^{2} u + \frac{\alpha}{2}  \left| u \right|^{4}u=0,
%-  \bigg( \left| \psi \right|^{2} - \frac{\alpha}{2}  \left| \psi \right|^{4}\bigg) \psi
\end{equation}
which describes a saturable Kerr-like nonlinearity through its truncated expansion at second-order in the normalized intensity $\left| u \right|^{2}$, with
$\alpha$ being an {\em external} free parameter that weights the quintic nonlinear response (the smaller $\alpha$, the weaker the saturation effect). 
Solitons of such system have been reported before. However, we reformulate the full problem from the beginning,
giving novel analytical formulas which prove convenient for the purpose of our analysis.
Solitons correspond to translationally invariant solutions of the  the form $u(z,x)=\sqrt{\rho(\theta)} \expÊ[i \phi(\theta) + i \beta z]$, where $\theta=x-v z$ 
and $\beta=g(\rho_0)$ is the nonlinear phase shift experienced by a plane-wave background with intensity
$\rho_0 \equiv |u_0|^2$, in the medium where the nonlinear refractive index varies with intensity $\rho=\left| u \right|^{2}$  according to the law 
$g(\rho)=- \rho + \alpha \rho^{2}/2$. 
%serve x l'Hamiltoniana
These solutions depend on two {\em internal} parameters which we choose, in analogy to general dark soliton solutions of the defocusing NLS equation, as $\rho_0$ (intensity background, which fixes also $\beta$) and $v$ (soliton velocity, which fixes also the darkness or brightness of the soliton).
Note that, here, the quintic term prevents the simple rescaling to $\rho_0=1$ without rescaling $\alpha$, and the velocity
complicates further the scenario, so we keep the three parameters free.
The modulus $\rho$ plays the role of equivalent "position", and obeys the standard Hamiltonian dynamics with 
"momentum" $p=\dot{\rho} \equiv d\rho/d\theta$,
%where the Hamiltonian $\mathcal{H}$ reads as:
\begin{eqnarray} 
\dot{p}=-\frac{\partial \mathcal{H}}{\partial \rho}, \;
\dot{\rho}=\frac{\partial \mathcal{H}}{ \partial p}; \;\;\; \mathcal{H}=\frac{p^2}{2}+V(\rho),\nonumber  \\
\label{eqH} \\
V=2\rho\bigg[ \frac{\alpha}{3}(\rho^3-\rho_0^3)-(\rho^2-\rho_0^2)+2k_0(\rho-\rho_0)-\frac{c_0^2}{\rho_0}\bigg].\nonumber 
\end{eqnarray}
Here $c_0=v\rho_0$ and $k_0=\frac{v^2}{2}+\rho_0-\frac{\alpha}{2}\rho_0^2$. 
Once $\rho(\theta)$ is obtained by solving Eqs. (\ref{eqH}), the phase profile $\phi(\theta)$ can be found by integrating the following equation:
\begin{equation}
\dot \phi = v \left( 1 - \frac{\rho_0}{\rho} \right). \label{phase}
\end{equation}
\subsection{Soliton solutions}
\begin{figure}[h!] 
\centering
\includegraphics[width=8cm]{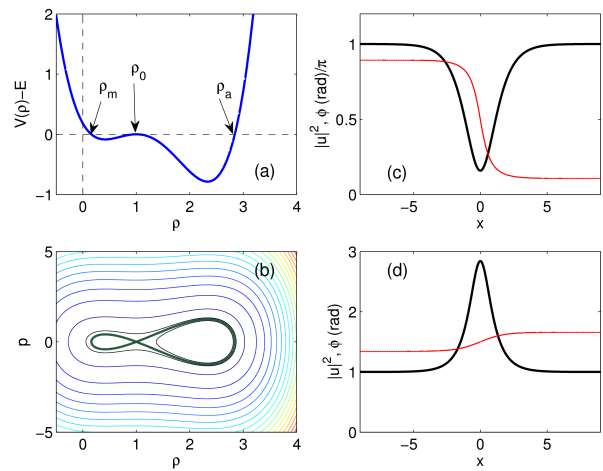}
\caption{(Color online) Dark-antidark pair sitting on the unit background $\rho_0=1$, for $\alpha=0.6$, $v=0.3$:
(a) potential $V(\rho)-E$ vs. $\rho$; (b) phase-space picture (contour lines of $\mathcal{H}$);
(c-d) Relative intensity (thick black solid line) and phase (thin red solid line) profiles of dark (c) and antidark (d)  solitons.} 
\label{f1}\end{figure}

Soliton solutions sitting on the plane-wave background $\rho=\rho_0$ correspond to homoclinic separatrix trajectories of Eqs. (\ref{eqH}),
characterized by the energy level $\mathcal{H}=E$ with $E=V(\rho_0)=-2c_0^2$. 
Such separatrices emanate from the saddle point $(\rho,p)=(\rho_0,0)$ of the Hamiltonian $\mathcal{H}(\rho,p)$.
For $\alpha \neq 0$, the potential $V(\rho)-E$ is a double well corresponding to a double-loop separatrix as shown in Fig. \ref{f1}(a,b).
Therefore one has, in general, a coexisting pair of dark and antidark solitons that corresponds to the motion along the
left well $\rho_m \le \rho \le \rho_0$ (dark solitons), and the right well $\rho_0 \le \rho \le \rho_a$ (antidark solitons), respectively.
Here $\rho_m \le \rho_0 \le \rho_a$ are the roots of $V(\rho)-E$ (explicit expressions of $\rho_m, \rho_a$ are reported in Appendix A).
In terms of such roots, we derive (see Appendix A) the following explicit solutions for dark solitons:
\begin{equation} \label{dark}
\rho_{d}(\theta) = \frac{\rho_m+r \rho_a \tanh^{2} \left[ w(\theta-\theta_{0}) \right]}{1+r \tanh^{2} \left[w(\theta-\theta_{0}) \right] },
\end{equation}
where $r=(\rho_0-\rho_m)/(\rho_a-\rho_0)$ and $w=\sqrt{\alpha (\rho_a-\rho_0)(\rho_0-\rho_m)/3}$ is the inverse soliton width,
while the minimum intensity (dip) is given by the root $\rho=\rho_m$.
Similarly for anti-dark solitons, we obtain
\begin{equation} \label{antidark}
\rho_{a}(\theta) = \frac{\rho_a+\frac{1}{r}\rho_m \tanh^{2} \left[ w(\theta-\theta_{0}) \right] }{1+ \frac{1}{r} \tanh^{2} \left[ w(\theta-\theta_{0}) \right]},
\end{equation}
where $r$ and $w$ are the same as for dark solitons.

%---------------------------- f1 -----------------------------------------
%---------------------------------------------------------------------------------
%---------------------------- f2 -----------------------------------------
\begin{figure}[h!] 
\centering
\includegraphics[width=6cm]{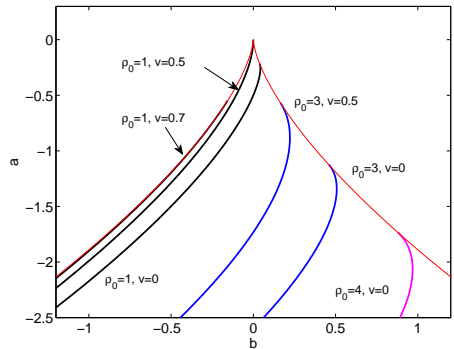}
\caption{(Color online) Cusp catastrophe picture for dark-antidark soliton pairs.
Each curve shows the evolution of the parameter $a$ and $b$ of the normal form potential 
$V(y)=y^4/4 + ay^2/2 + b y$, calculated for a dark-antidark soliton pair with fixed internal parameters $v$ and $\rho_0$,
and $\alpha$ changing from zero up to its critical value $\alpha_c$, where all the curves arrive
tangentially on the cusp curve (Eq. (\ref{cusp}), red solid line) that bounds the soliton existence domain.
} 
\label{f2}\end{figure} 
%---------------------------------------------------------------------------------
%---------------------------- f3 -----------------------------------------
\begin{figure}[h!] 
\centering
\includegraphics[width=8cm]{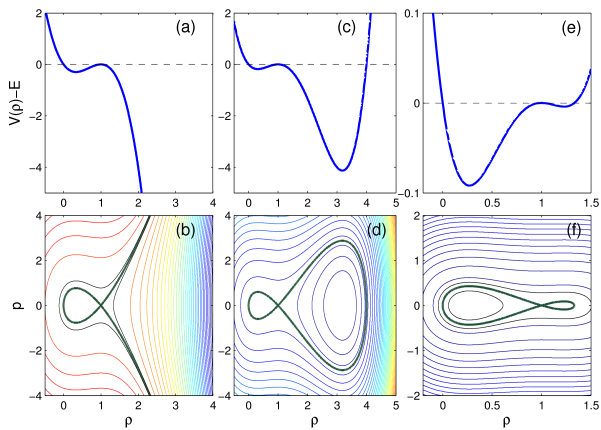}
\caption{(Color online) Potential and phase-space picture for $v=0$, $\rho_0 =1$, and different values of $\alpha$:
(a-b) $\alpha=0$ (ideal Kerr case); (c-d) $\alpha=0.5$;  (e-f) $\alpha=0.9$ (close to critical value $\alpha_c=1$,
where the right branch of the separatrix becomes vanishingly small).
} 
\label{f3}\end{figure} 
%---------------------------------------------------------------------------------
The expressions in Eqs. (\ref{dark})-(\ref{antidark}) allow us to obtain, by integrating Eq. (\ref{phase}),  the nonlinear phase associated to the two soliton families. 
We obtain for dark and antidark solitons, respectively
\begin{eqnarray} 
\phi_{d}=-\frac{v}{w} s \tan^{-1} \left( \sqrt{ \frac{ r \rho_a}{\rho_m}  } \tanh [w (\theta - \theta_0 )]\right) + \phi_0,\label{nlph1} \\
\phi_{a}=\frac{v}{w} s \tan^{-1} \left( \sqrt{ \frac{\rho_m}{r \rho_a} } \tanh [w (\theta - \theta_0 )]\right)+ \phi_0,\label{nlph2}
\end{eqnarray} 
where $s \equiv \sqrt{ \frac{(\rho_a-\rho_0)(\rho_0-\rho_m)}{\rho_a \rho_m} }$. From  Eqs. (\ref{nlph1}-\ref{nlph2}) one can easily calculate the
phase jump $\Delta \Phi=\Phi(+\infty)-\Phi(-\infty)$ across the soliton. 

Interestingly, the existence domain of the soliton pairs can be described with the aid of catastrophe theory \cite{gilmore},
already applied to characterize dark-antidark soliton pairs in a different context (gap soliton theory \cite{Conti01}).
In fact, the quartic potential $V$ in Eqs. (\ref{eqH}) belongs to the $A_+$ family, and gives rise to the so called \emph{cusp} catastrophe. 
According to this picture the potential $V(\rho)$ in Eq. (\ref{eqH}), which is of the general form $V(\rho)=c_4 \rho^4 + c_3 \rho^3 + c_2 \rho^2 + c_1 \rho$,
can be cast into the canonical form \cite{gilmore} $V(y)=y^4/4 + ay^2/2 + b y$, 
by means of the change of variable $\rho=(4c_4)^{-1/4} y -c_3/(4c_4)$.
Then, in the control parameter plane $(a,b)$ 
(explicit expressions of $a$ and $b$ as a function of $\alpha, \rho_0, v$ are cumbersome but can be easily derived),
solitons exist in the inner region bounded by the curve (so called bifurcation set  \cite{gilmore}) 
\begin{equation} \label{cusp}
\left( \frac{a}{3} \right)^3 + \left( \frac{b}{2} \right)^2=0,
\end{equation}
shown in Fig. \ref{f2}.
Such curve marks the values where the critical points ($\partial_y V=0$) of the potential
become doubly degenerate ($\partial_y^2 V =0$), 
and exhibits the characteristic shape of a cusp in the origin (three-fold degenerate point, $\partial_y^3 V =0$). 
In terms of the original parameter $\rho_0, v, \alpha$, the existence condition requires $\alpha \le \alpha_c$, 
with the following critical value of the quintic coefficient $\alpha_c$:
\begin{equation} \label{ex}
\alpha_c=\frac{1}{\rho_0}\bigg( 1-\frac{v^2}{\rho_0}\bigg).
\end{equation}
Taking fixed internal parameters $\rho_0$ and $v$, while changing $\alpha$ continuosly, makes the control parameters $a$ and $b$
calculated for the soliton to span a smooth curve in the control parameter plane $(a,b)$, 
until at $\alpha=\alpha_c$, the curve hits (arriving tangentially) the boundary set by the cusp curve [Eq. (\ref{cusp})]. 
Different values of $\rho_0$ and $v$ result into different control curves, as displayed in Fig. \ref{f2}. 
We point out that a similar behavior occurs by varying $\rho_0$ or $v$, keeping the other two parameters fixed.
In particular, in the latter case, the existence domain turns out to be $-v_c \le v \le v_c$, 
with the cut-off velocity $v_c=\sqrt{\rho_0-\alpha \rho_0^2}$, obtained by expressing  Eq. (\ref{ex}) in terms of $v=v_c$.
%---------------------------- f4: darkness -----------------------------------------
\begin{figure}[h!] 
%\centering
\hspace{-1cm}
\includegraphics[width=9cm]{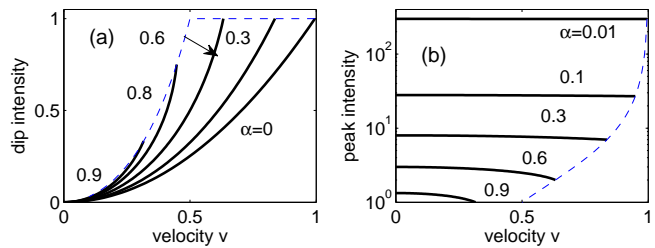}
\caption{(Color online) (a) dip intensity (darkness) of dark solitons and (b) maximum intensity of antidark solitons 
as a function of velocity $v$ for different values of $\alpha$. Here the background is $\rho_0=1$.
The dashed curves represent the existence threshold set by Eq. (\protect{\ref{ex}}).
Note the vertical log scale in (b).} 
\label{f4}\end{figure} 
%------------------------------------------------------------------------------

As an example, we show in Fig. \ref{f3} how the typical phase plane (potential) changes when $\alpha$ is varied between zero
and the critical value $\alpha_c$. In this case we choose $\rho_0=1$ and still solitons, viz. $v=0$, yielding $\rho_m=0$ 
which means that the dark soliton is black regardless of the value of $\alpha$, while the antidark is characterized by a peak intensity
$\rho_a=3/\alpha - 2\rho_0$. In the limit $\alpha=0$ shown in Fig. \ref{f3}(a,b), which represents the ideal Kerr case, the potential is cubic, 
and the separatrix has only one branch corresponding to the  well-known black soliton solution ($\rho=\tanh^2(x)$) of the NLS equation,
whereas for $\rho>1$ the motion is unbounded and no coexisting antidark solutions do exist. In fact the unbounded motion
for $\rho>1$ can be thought of as the motion in the right well that, however, becomes infinitely deep [$V(\theta=\infty) \rightarrow -\infty)$] and wide (since $\rho_a \rightarrow \infty$).
Viceversa, as $\alpha$ grows from zero, the behavior of the potential at $\theta=\infty$ is inverted, and the right well becomes finite,
allowing for a eight-shaped separatrix corresponding to the dark-antidark pair [see Fig. \ref{f3}(c-d)].  
For small values of $\alpha$ the antidark soliton has high peak intensity $\rho_a$ above the background $\rho_0$, 
which, however, decreases as the saturation parameter $\alpha$ increases. For $\alpha$ approaching its critical value $\alpha_c$
antidark solitons become shallow [see Fig. \ref{f3}(e-f)], until they reduce to the plane wave exactly at $\alpha=\alpha_c$, 
where $\rho_a \rightarrow \rho_0$. 
%---------------------------- f5: bistability -----------------------------------------
\begin{figure}[h!] 
%\centering
\hspace{-1cm}
\includegraphics[width=9cm]{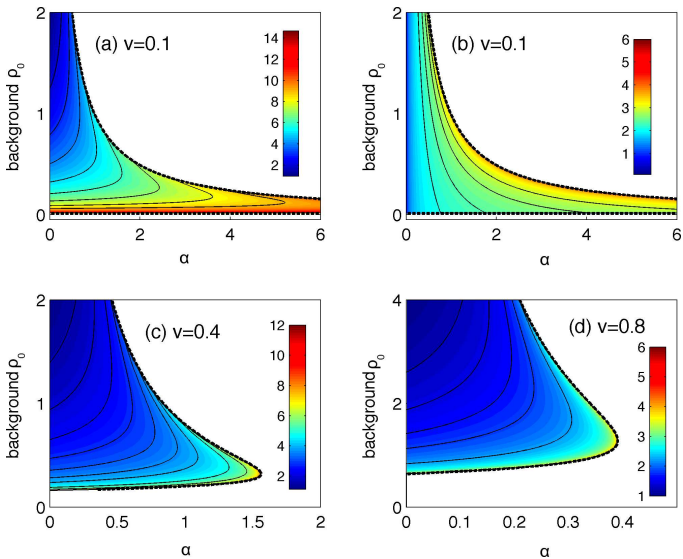}
\caption{(Color online) Color level plot of FWHM [$\theta_{FWHM}$ in Eq. (\protect{\ref{fwhm}})]
in the parameter plane $(\alpha, \rho_0)$ for $v=0.1$, dark (a) and antidark (b) solitons.
(c,d) dark solitons  with $v=0.4$ (c), and $v=0.8$ (d).
The dashed lines give the critical condition $\alpha_c(\rho_0,v)$.} 
\label{f5}\end{figure} 
%------------------------------------------------------------------------------
The behavior of dark-antidark solitons with velocity exhibits intriguing features, which can be gathered
by plotting the minimum (dip) intensity $\rho_m$ of dark solitons (the larger $\rho_m$, the lower the darkness)
and the peak intensity $\rho_a$ of antidark solitons (the higher $\rho_a$, the brighter the antidark)
versus $v$ at constant $\rho_0$, for different values of $\alpha$, as displayed in Fig. \ref{f4}.
For non-zero but small velocities, the picture remains qualitatively unchanged with respect to the case $v=0$, 
in the sense that the peak intensity $\rho_a$ of antidark solitons decreases continuously from infinity (at $\alpha=0$) 
to $\rho_a=\rho_0$ at the critical value $\alpha_c$, such that the solitons become infinitely shallow (i.e. they reduce to a pure plane wave). 
In this case, however, the dip intensity of dark solitons is no longer zero, i.e. they becomes gray solitons with darkness $\rho_0-\rho_m$. 
The darkness decreases for growing velocities $v$ up to a minimum value at the bound velocity $v_c=\sqrt{\rho_0-\alpha \rho_0^2}$ 
(obtained by solving Eq. (\ref{ex}) with respect to $v$ for fixed $\alpha$).
This is clearly shown in Fig. \ref{f4}, where we summarize the result for a fixed background $\rho_0=1$.
Note from Fig. \ref{f4}(a) that Kerr (NLS) dark solitons ($\alpha=0$) are always darker than the corresponding CQNLS solitons of the same velocity
(the curve $\rho_m(v)$ for $\alpha=0$ is always below the other curves relative to $\alpha \neq 0$), 
and have also a larger phase jump $\Delta \Phi$ than their CQNLS counterparts.

Interestingly, however, as the velocity grows large enough (above $v=0.5$ in Fig. \ref{f4}), 
it turns out that dark solitons can become infinitely shallow at the cut-off condition for their existence 
($\rho_a \rightarrow \rho_0$ and hence darkness tends to zero).
Conversely, under the same conditions, antidark solitons cease to become infinitely shallow,  rather 
reaching a finite minimum peak intensity at the cut-off condition for their existence [see Fig. \ref{f4}(b)].
From Fig. \ref{f4}(b) it is also clear that, for small $\alpha$ the brightness of antidark solitons is nearly independent
on the velocity. This change of behavior at the cut-off condition for the existence (from infinitely shallow antidark to infinitely shallow
dark solitons) depends the background $\rho_0$. It can be shown to occur at the value of $\alpha_0=3/(4\rho_0)$,
in correspondence of the velocity $v_0=\sqrt{\rho_0}/2$ (in Fig. \ref{f4}, $\alpha_0=0.75$ and $v_0=0.5$).

The coexistence of dark and antidark solitons constitute a bistable mechanism such that two different solutions
exist with same parameters ($\rho_0$ and $v$) and different renormalized invariants $M, H, P$ (see Appendix A for their definition).
However these soliton families can be bistable also according to the definition by Gatz and Herrmann, 
i.e. for fixed $\alpha$ different solutions of the same width can exist although they sit on a different background $\rho_0$. 
This type of bistability was investigated for still ($v=0$) dark solitons \cite{Herrmann92}.
In order to generalize this result to the full family (any $v$, and antidark case), we have calculated the FWHM as
\begin{eqnarray} \label{fwhm}
\theta_{FWHM}=\frac{2}{w} \tanh^{-1} \sqrt{f(\rho_0, \rho_m, \rho_a)},
\end{eqnarray}
where $f(\rho_0, \rho_m, \rho_a)= \frac{\rho_a-\rho_0}{2\rho_a-\rho_0 - \rho_m}$ for dark solitons 
(FWHM taken at half the intensity between $\rho_0$ and the dip $\rho_m$)  
and $f(\rho_0, \rho_m, \rho_a)= \frac{\rho_0-\rho_m}{\rho_a+\rho_0 - 2\rho_m}$ for antidark solitons
(FWHM at half the intensity between $\rho_0$ and the peak $\rho_a$), respectively.
The results obtained by mapping $\theta_{FWHM}$ in the plane ($\alpha, \rho_0$) are summarized in Fig. \ref{f5}.
As shown, dark solitons exhibit bistability for sufficiently large $\alpha$, regardless of their velocity. 
However the range of values of $\alpha$ where bistability occurs is greatly reduced for large velocities [see Fig. \ref{f5}(c)]. 
Conversely, as shown in Fig. \ref{f5}(d),  antidark solitons are never bistable in the same sense (no folding of the level curves of $\theta_{FWHM}$ is ever observed at any velocity).
%---------------------------- f6 : M vs v -----------------------------------------
\begin{figure}[h!] 
\centering
\includegraphics[width=8cm]{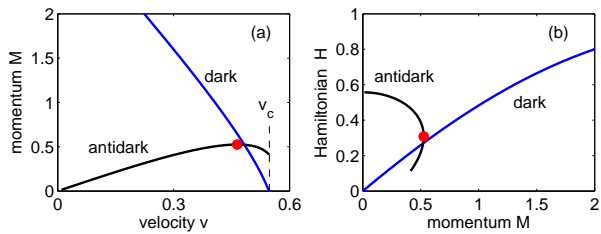}
\caption{(Color online) 
(a) Renormalized momentum $M$ vs. soliton velocity $v$, and (b) Hamiltonian $H(v)$ vs. $M(v)$, for dark and antidark solitons
with fixed $\rho_0=1$ and $\alpha=0.7$. The red bullets mark the marginal condition $\partial_v M=0$ for antidark solitons.} 
\label{inv}\end{figure} 
%-----------------------------------------------------------------------------------
%---------------------------- f7 : stab map -----------------------------------------
\begin{figure}[h!] 
\centering
\includegraphics[width=9cm]{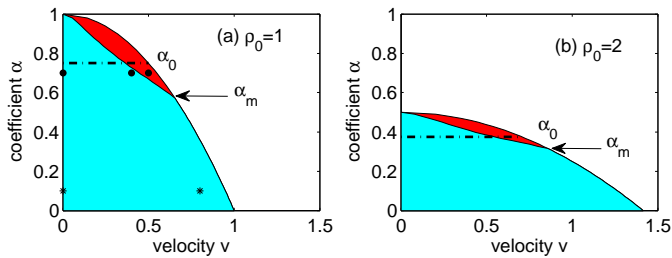}
\caption{(Color online) Stability maps for antidark solitons in the parameter plane $(v,\alpha)$ for (a) $\rho_0=1$;  (a) $\rho_0=2$.
The light (cyan) and dark (red) shaded domains correspond to unstable and stable solutions, respectively.
The dot-dashed line sets the value $\alpha_0$ below which antidark solitons have finite brightness at cut-off.
The dynamics illustrated in Fig. 8 and Figs. 9-10 are relative to the sampled values marked by
bullets and stars in (a), respectively.} 
\label{map}\end{figure} 
%-----------------------------------------------------------------------------------
\section{INSTABILITY SCENARIOS}

Having characterized the features of dark-antidark soliton pairs, we discuss their stability.
We proceed by applying a known stability criterium according to which the stability is related to the derivative 
of the invariant momentum $M$ (see Appendix A)
of the soliton against its velocity $v$. The marginal condition 
\begin{equation} \label{marginal}
\frac{\partial M}{\partial v} =0,
\end{equation}
separates stable solutions ($\partial_v  M  <0$) from unstable ones ($ \partial_v  M >0$).
Such criterium, proved by Barashenkov \cite{PhysRevLett.77.1193} for dark solitons of the generalized NLS equation
is found to account also for the instabilities of antidark solitons. We emphasize, however, that such criterium
accounts for real eigenvalues crossing into the right-half plane (usually bifurcating from zero eigenvalues associated
with neutral modes, i.e. symmetries of the model). Other instability mechanisms such as oscillatory instabilities resulting from edge bifurcations
leading to pairs of complex conjugate eigenvalues entering the right-half plane, need an independent characterization.
However, we have found (numerically) no evidence for instabilities of such kind. Moreover, the condition for the existence
of solitons [Eq. (\ref{ex})] turns out to coincide with the region where the background plane-wave is modulationally stable.
Therefore the condition (\ref{marginal}) turns out to be an exhaustive criterium to assess the linear stability 
of all the solutions presented so far.

The calculation of the function $M(v)$ for dark soliton family shows that such function has always a negative slope. 
Therefore the whole family of dark solitons is stable in its existence domain, 
a conclusion that is fully supported by our numerical simulation of the propagation.

Viceversa instabilities take place for antidark solitons. In particular, for small $\alpha$, it turns out that 
they are always unstable in the whole domain of existence since $M(v)$ exhibits always positive slope.
Conversely at sufficiently large $\alpha$, the momentum $M(v)$ changes its slope near the cut-off value for existence $v_c$.
This is shown in Fig. \ref{inv}, where we compare the momentum $M(v)$ for dark and antidark
solitons at $\alpha=0.7$ and $\rho_0=1$. Note that the change of slope of the momentum means
that two antidark solutions with same momentum and different velocity exist.
These two solutions differ by their Hamiltonian $H$ (see Appendix for its expression).
Indeed, as shown in Fig. \ref{inv}(b) the Hamiltonian $H$ as a function of $M$ is folded,
showing an upper (unstable) branch and a lower (stable) branch.
%---------------------------- inst large alfa -----------------------------------------
\begin{figure}[h!] 
\centering
\includegraphics[width=7cm]{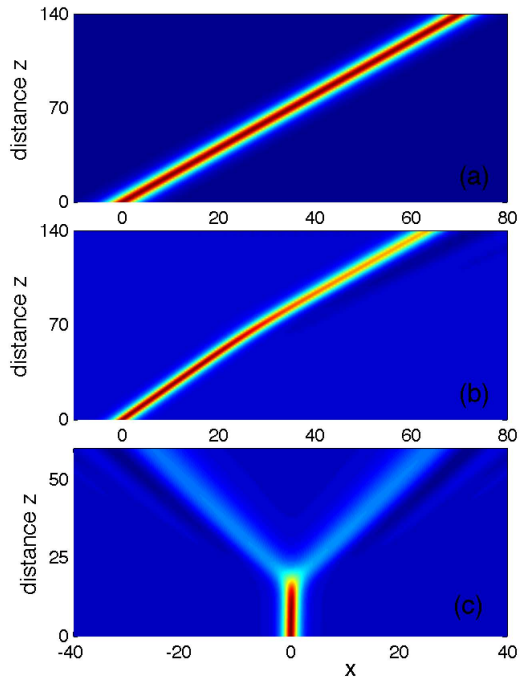}
\caption{(Color online) Dynamics of antidark solitons with $\rho_0=1$, $\alpha=0.7$, exhibiting different behavior depending
on the initial velocity $v$ [see bullets in Fig. 7(a)]: (a) $v=0.5$, stable propagation;
(b) $v=0.4$, decay into a nearby stable antidark soliton; 
(c) $v=0$, decay into a pair of antidark shallow solitons with opposite velocities.} 
\label{inst}\end{figure} 
%-----------------------------------------------------------------------------------
%---------------------------- dsw 1 -----------------------------------------
\begin{figure}[h!] 
\centering
\includegraphics[width=7cm]{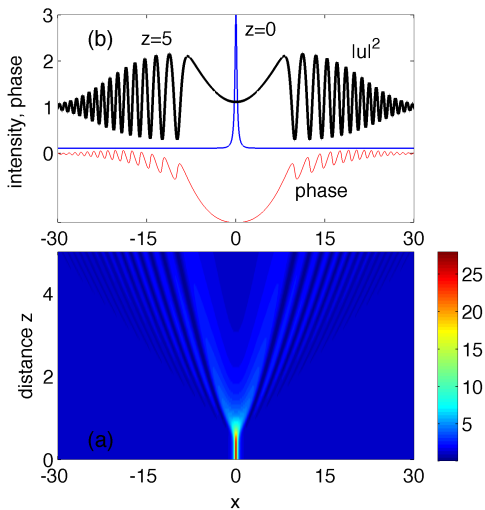}
\caption{(Color online) 
Decay of an unstable antidark solitons ($\rho_0=1$, $v=0$, $\alpha=0.1$) into two symmetric dispersive shock fans (trains of dark solitons):
(a) color level plot of the intensity; (b) snapshot of intensity (black solid line) and phase (red solid line) at $z=5$. For comparison
the input intensity (renormalized to the maximum of the plot) is shown (solid blue line).} 
\label{dsw1}\end{figure} 
%-----------------------------------------------------------------------------------
%---------------------------- dsw 2 -----------------------------------------
\begin{figure}[h!] 
\centering
\includegraphics[width=7cm]{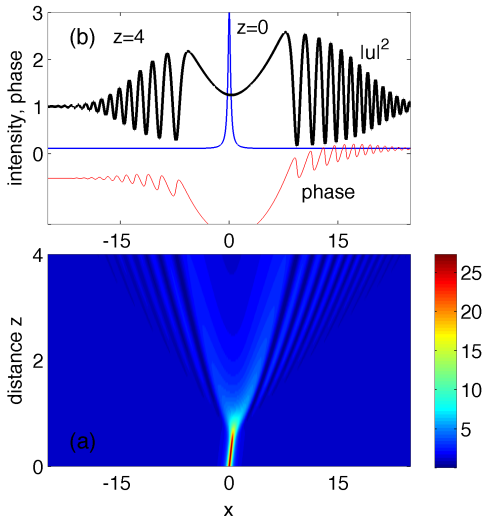}
\caption{(Color online) As in Fig. 7 for $v=0.8$.} 
\label{dsw2}\end{figure} 
%-----------------------------------------------------------------------------------

The velocity which gives the marginal stability ($\partial_v M=0$) clearly depends on $\rho_0$ and $\alpha$.
The results of Fig. \ref{inv} could be repeated for different values of $\rho_0$ and $\alpha$,
and summarized by drawing stability maps in the plane $v, \alpha$ at constant $\rho_0$.
These are displayed in Fig. \ref{map} for two different values of the background $\rho_0$. 
As shown a relatively small island of stability is found in the vicinity of the boundary for existence, where they have small brightness.
In particular stability requires $\alpha > \alpha_m$, where $\alpha_m$ corresponds to the vertex of the stability island [see Fig. \ref{map}].
In order to test the validity of the marginal stability condition we have made extensive simulations
performed by means of the well-known split-step method.
Such simulations confirm indeed that $\partial_v M=0$ gives the threshold for stability also for antidark solitons.
% rewretten sentence in the revised version
The numerics also reveal two basic mechanisms of instability, which are illustrated below by means of
numerical runs performed by launching the exact soliton profile, 
while the perturbation arises from intrinsic roundoff and discretization errors.\\
\indent The first scenario is valid for relatively large $\alpha$, such that a stable range of velocities exists.
As an example for illustration purpose, we have chosen $\alpha=0.7$, $\rho_0=1$, which yield
a range of stable velocities $v=(0.4644,0.5477)$ [see Fig. \ref{inv}(a)]. 
The dynamics for three different values of $v$, marked by bullets in the map of Fig. \ref{map}(a), is displayed in Fig. \ref{inst}. 
As shown in Fig. \ref{inst}(a), for $v=0.5$ the antidark soliton propagates in stable fashion. 
Conversely, when $v$ is decreased just below the threshold for instability [$v=0.4$ in Fig. \ref{inst}(b)], 
the initial soliton decays into a different antidark soliton of higher velocity, which is stable. 
However, for lower velocities, such that the soliton lies far from the threshold for stability, the decay into a pair of stable antidark soliton
is observed, a process which is accompanied by radiation. Note that the instabilities are extremely long-range in both cases,
i.e. they become manifest only after tens of diffraction lengths. As shown in Fig. \ref{inst}(c), the two antidark solitons are symmetric in the limit
$v=0$ in such a way that the initial zero momentum is conserved, otherwise they appear to be asymmetric (case not shown). 
The scenario illustrated in Fig. \ref{inst} holds also for different values of $\alpha$, providing $\alpha>\alpha_m$,
i.e. a stable range of velocities exists. It is interesting to note that the value $\rho_0=3/(4\rho_0)$, which discriminates
the fact that antidark solitons can be or not infinitely shallow at cut-off, lies in the region of stability for any $\rho_0$
(see dot-dashed lines in Fig. \ref{map}). Therefore antidark solitons with arbitrarily small brightness that exist  for $\alpha> \alpha_0$ around cut-off,
are always stable, whereas antidark solitons which have finite brightness at cut-off can propagate stably only for $\alpha>\alpha_m$.\\
\indent The scenario discussed above change qualitatively when $\alpha<\alpha_m$, where the island of stability shrinks to zero.
In this case the decay instability towards other antidark solitons is forbidden because no stable solutions exist.
In this case, the only stable solutions are dark solitons and therefore the decay instability occurs towards these solutions,
even though their shape differ dramatically (indeed being "opposite") from the input antidark shape.
Importantly when $\alpha$ is small (weak saturation) the antidark solitons possess large amplitude and large power $P_a$ (or number of
particles, see Appendix for its definition). Under these conditions we have found that the decay instability of the antidark
leads to a DSW (see Refs. \cite{Hoefer06,Wan07,PhysRevLett.102.083902,PhysRevA.83.053846} and references therein), 
i.e. an expanding region filled with fast oscillations which behave asympotically as solitons. 
In the present case, starting with a zero-velocity antidark soliton, the decay instability leads to two symmetric DSW fans, 
as displayed in Fig. \ref{dsw1} for $\alpha=0.1$. In each of the two fans, the inner edge is set by the darkest and slowest soliton,
whereas on the outer edge the fan is linked to the plane wave through a train of solitons with progressively decreasing darkness and increasing velocity,
which become denser as the background is approached. Although this behavior is reminiscent of that ruled
by the integrable NLS equation (limit $\alpha=0$) in the semiclassical regime (i.e., nonlinearity much stronger than diffraction/dispersion)
under excitation of e.g. a gaussian on pedestal \cite{Wan07}, it must be emphasized that this is, to the best of our knowledge,
the first example where the dispersive shock waves occurs directly through the decay of an unstable solitary wave of the system.
Further characterization of the shock fan need to develop the Whitham modulational theory for the system, which is beyond the scope
of this paper. However, simple arguments based on the features of solitons as shown in Fig. \ref{f4}, let us predict that the fan
ruled by Eq. (1) is narrower as compared to the one ruled by the integrable NLS equation under the same excitation.
This is due to the fact that the soliton with vanishingly small darkness, constituting the outer edge of the fan, 
correspond to progressively reduced velocity as the quintic nonlinearity grows (i.e., $\alpha$ increases), as clearly shown in Fig. \ref{f4}(a).\\
\indent The decay scenario shown in Fig. \ref{dsw1} does not change when starting with an antidark soliton with non-zero velocity ($v,M \neq 0$),
except that the two shock fans become asymmetric (the higher the velocity, the higher the asymmetry), 
in both the number of solitons and the velocity of the darkest soliton (inner edge of the fan), as shown in Fig. \ref{dsw2}. 
This asymmetry is definitely expected, based on the fact that a symmetric configuration would have $M=0$ and hence would lead to violation of momentum conservation. Moreover the asymmetric development of the shock is analogous to DSW generated in the integrable NLS limit when starting
from a gray beam with non-zero velocity and momentum \cite{PhysRevA.83.053846}.\\
\indent Finally we point out that, for larger $\alpha$, yet with $\alpha < \alpha_c$, the scenario remains qualitatively unchanged though
the number of dark solitons generated in the decay of the antidark soliton decreases.\\ 
\indent Though the aim of this paper was the full characterization of solitary waves sitting on a finite background,
before concluding, we point out that such solitons can also coexist with bright solitary waves with zero pedestal.
Seeking for bright solitons of the form $u(x,z)=\sqrt{\rho(x)} \exp(i \beta z)$, a simple calculation shows that
the peak intensity of these solutions turns out to be $\rho_{a} =3(1 + \sqrt{1+ 8 \beta \alpha/3})/(2 \alpha)$.
The existence domain of such solutions include arbitrarily small $\alpha$, and hence they coexist with dark-antidark pairs. 
Given the defocusing nature of the leading order (cubic) nonlinearity this might appear surprising. 
However, it is not difficult to understand that such bright solitons are sustained entirely by the quintic focusing nonlinearity
since $\rho_{a}$ diverges in the limit $\alpha \rightarrow 0$, where the quintic nonlinearity vanishes indeed.
Without deepening the study of the bright case, already investigated in the framework of the CQNLS equation, 
we expect them to be unstable at least in the limit of small $\alpha$, where they do not affect the decay dynamics
of antidark solitons or the stable dynamics of dark solitons discussed above.
 
\section{Conclusions}
In summary we have discussed the main properties of solitary solutions with finite background
of the CQNLS equation with focusing quintic term.
Bistability and instabilities have been studied for the full family of solutions obtained in new analytical form,
and parametrized by the intensity of the background wave and the velocity.
We have found that the solutions exhibit a non-trivial behavior against the velocity such that,
depending on the value of the quintic nonlinearity, either dark or antidark solitons, become infinitely shallow at their bound for existence.
Furthermore dark solitons are bistable for any velocity, whereas antidark are never bistable.
Finally dark solitons are stable against weak perturbations while antidark are mostly unstable, exhibiting
different mechanisms of instability. In particular, a novel mechanism involving the decay of an antidark soliton into a dispersive shock wave has been characterized. Further work will be devoted to assess how the quintic nonlinearity affects the formation and dynamics of dispersive shock waves
which develop from more general (non-solitary) inputs. Potentially also the collapse, i.e. blow-up at a finite distance
(which is known to occur even in the 1+1D case that we dealt with, owing to the high order of the focusing term \cite{Glassey77,Pathria96}) 
could play a substantial role that needs further investigation.\\
\indent From the experimental point of view, the most natural setting for testing these results is 
the study of paraxial beam evolution in nonlinear optics of centrosymmetric media, 
where the quintic term accounts for the saturation of the Kerr nonlinear index $n_2$ ($n_2<0$, defocusing media),
usually quantified in terms of the high-order nonlinear index $n_4$ \cite{Herrmann92},  
the overall index change being $\Delta n  = -|n_2| I + n_4 I^2$, where $I$ is the optical intensity.
While the nonlinear indexes $n_2, n_4$ (or equivalently the nonlinear susceptibilities)
depends solely on material properties and can be accurately characterized by means of consolidated techniques \cite{Wise05,Zhang07}, 
the normalized coefficient $\alpha$ used throughout the paper turns out to depend on the input intensity as well, 
and hence the impact of the quintic nonlinearity can be tuned by changing the optical power \cite{PhysRevA.81.053831}. 
An other area where the predictions based on the present model can be relevant and can lead to experimental test,
is the dynamics of ultracold atoms (Bose-Einstein condensates), 
where the quintic term arises from higher-order (three-body) atom interactions \cite{Abdullaev01},
and tuning of the nonlinearities can be achieved by means of Feshback resonances.

\vspace{1cm}
\section{Appendix A}

Analytical solutions can be worked out from the equation which follows directly from Eq. (\ref{eqH})
\begin{equation} \label{eqrho}
\dot \rho=   \sqrt{\frac{4 \alpha}{3}} \sqrt{Q(\rho)},
\end{equation}
where $Q(\rho)= \frac{3}{4 \alpha} \left[ E - V(\rho) \right]$. The polynomial $Q(\rho)$ can be expressed in terms
of its ordered roots $\rho_m$, $\rho_0$  (double root), $\rho_a$, with ordering $\rho_m \le \rho_0 \le \rho_a$,
as $Q(\rho)=(\rho- \rho_a)(\rho-\rho_m)(\rho-\rho_0)^2$.
From Eq. (\ref{eqrho}) one obtains the following quadrature integral,
\begin{eqnarray} \label{quad}
\int^{\rho(\theta)}_{\rho(\theta_0)} \frac{d \rho}{\sqrt{Q(\rho)} } = \sqrt{\frac{4 \alpha}{3}} \int_{\theta_0}^{\theta} d \theta,
\end{eqnarray}
where $\rho(\theta_0)=\rho_m$ or $\rho(\theta_0)=\rho_a$ for dark or antidark solitons, respectively.
The integral (\ref{quad}) gives, upon inversion, the solutions given in the text.
Explicitly, the extremal roots $\rho_m, \rho_a$ are expressed by:
\begin{align}
\rho_m=\frac{3-2 \alpha \rho_0-\sqrt{(3-2 \alpha \rho_0)^2-12 v^2 \alpha}}{2 \alpha },\nonumber\\
\rho_a=\frac{3-2 \alpha \rho_0+\sqrt{(3-2 \alpha \rho_0)^2-12 v^2 \alpha}}{2 \alpha }.
\end{align}
Note that, in the limit $v=0$, the smaller root $\rho_m$ vanishes, and as a consequence the dark soliton
becomes a black soliton.
%\begin{equation}
%\rho^{\pm} - \rho_0=\frac{2v^2 - 2\rho_0\left( 1- \alpha \right)}{1-\frac{4}{3} \alpha \rho_0 
%\pm \sqrt{\left( 1- \frac{4}{3} \alpha \rho_0 \right)^2 + \frac{2 \alpha}{3} \left( 2\rho_0\left( 1 - \alpha \right) - 2v^2 \right) }}
%\end{equation}

Once the soliton solutions are known one can easily calculate the renormalized invariants (momentum, Hamiltonian)
which are defined as follows
\begin{equation}
M= \frac{i}{2} \int_{-\infty}^{+\infty} (u_x^* u - u_x u^*)\left(1-\frac{\rho_0}{\rho} \right) \,dx,
\end{equation}
\begin{equation}
H=  \int_{-\infty}^{+\infty}  \left[\frac{|u_x|^2}{2}  + \int_{\rho_0}^{\rho(x)}   [g(\rho_0) - g(\rho)] d\rho \right] dx ,
\end{equation}
and the power or number of particles, $P_a$ (antidark) and $P_d$ (dark):
\begin{eqnarray}
P_a=\int_{-\infty}^{+\infty} |u|^2 - \rho_0  \,dx;\;P_d=\int_{-\infty}^{+\infty}  \rho_0 - |u|^2 \,dx
%P=\int_{-\infty}^{+\infty} dx \; |u|^2 - \rho_0 \;\; (antidark)
\end{eqnarray}
The quantities $M$ and $H$ are those employed in the text to assess the stability of soliton solutions.

\section{Acknowledgements}
S. T. acknowledges funding by italian Ministero dell'Istruzione, dell'Universit\`{a} e della Ricerca (MIUR),
in the framework of PRIN 2009 project, No. 2009P3K72Z.
%A. F. acknowledges funding by Award No. KUK-F1-024-21 (2009/2012) made by King Abdullah University of Science and Technology (KAUST).

%\bibliography{sol35} 

%
%-----------------------
\end{document}